\documentclass[12pt]{iopart}
\usepackage{graphicx}
\usepackage{afterpage}
\usepackage{lscape}
\usepackage{fancyhdr}
\usepackage{colortbl}
\usepackage{lineno}
\usepackage{latexsym}

\begin{document}

\def\Journal#1#2#3#4{{#1} {\bf #2}, #3 (#4)}

\def\NCA{Nuovo Cimento}
\def\NIM{Nucl. Instr. Meth.}
\def\NIMA{{Nucl. Instr. Meth.} A}
\def\NPB{{Nucl. Phys.} B}
\def\NPA{{Nucl. Phys.} A}
\def\PLB{{Phys. Lett.}  B}
\def\PRL{Phys. Rev. Lett.}
\def\PRC{{Phys. Rev.} C}
\def\PRD{{Phys. Rev.} D}
\def\ZPC{{Z. Phys.} C}
\def\JPG{{J. Phys.} G}
\def\CPC{Comput. Phys. Commun.}
\def\EPJ{{Eur. Phys. J.} C}

\title[Production and energy loss of strange and heavy quarks]{Production and energy loss of strange and heavy quarks}

\author{Lijuan Ruan (for the STAR Collaboration)}

\address{Physics Department, Brookhaven National Laboratory, Upton, New York, NY 11973, USA}
\ead{ruanlj@rcf.rhic.bnl.gov}
\begin{abstract}

Data taken over the last several years have demonstrated that RHIC
has created a hot, dense medium with partonic degrees of freedom.
Identified particle spectra at high transverse momentum ($p_T$)
and heavy flavor are thought to be well calibrated probes thus
serve as ideal tools to study the properties of the medium. We
present $p_T$ distributions of particle ratios in p+p collisions
from STAR experiment to understand the particle production
mechanisms. These measurements will also constrain fragmentation
functions in hadron-hardon collisions. In heavy ion collsions, we
highlight: 1) recent measurements of strange hadrons and heavy
flavor decay electrons up to high $p_T$ to study jet interaction
with the medium and explore partonic energy loss mechanisms; and
2) $\Upsilon$ and high $p_T$ $J/\psi$ measurements to study the
effect of color screening and other possible production
mechanisms.

\end{abstract}

\section{Introduction}

The physics goal at RHIC is to study and identify the properties
of matter with partonic degrees of freedom~\cite{starwhitepaper}.
Hard probes such as identified particles at high $p_T$, heavy
flavor and jets are thought to be well calibrated probes since
they are calculable in the perturbative Quantum Chromodynamics
(pQCD) framework~\cite{pQCD}. The single inclusive hadron
production in p+p collisions at $p_T\!>\!2$ GeV/$c$ is described
by the convolution of parton distribution functions, parton-parton
interaction cross sections and fragmentation functions (FFs). In
Au+Au collisions, particle production at $p_T\!>\!6$ GeV/$c$ can
be described by the fragmentation process with jet energy loss.
When jets traverse the hot and dense medium, they lose energy
through gluon radiation and/or colliding elastically with
surrounding partons~\cite{jetquench,starhighpt,rhicotherhighpt}.
This leads to a softening of hadron spectra at high $p_T$. The
softening can be characterized through nuclear modification factor
($R_{AA}$ ($R_{CP}$)), the spectrum in central Au+Au collisions
divided by that in p+p (peripheral Au+Au) collisions, scaled by
the number of underlying binary nucleon-nucleon inelastic
collisions ($N_{bin}$). The amount of energy loss can be
calculated in QCD and is expected to be different for light
quarks, gluons and heavy quarks~\cite{xinnian:98,deadcone}.
Through the comparison between $R_{AA}$ or $R_{CP}$ measurements
with theoretical calculations, medium properties such as gluon
density and transport coefficient can be derived.

To further understand energy loss mechanisms and medium
properties, we measured nuclear modification factors for protons,
pions and non-photonic electrons at STAR to test color charge and
flavor dependence of energy loss. For example, gluons carry
different Casmir factor from quarks. The coupling of the gluon to
the medium is stronger than the coupling of quark to medium thus
gluons are expected to lose more energy than quarks when
traversing the medium. At RHIC energy, gluon jet contribution to
protons is expected to be significantly larger than to pions at
high $p_T$~\cite{xinnian:98,AKK,ppdAuPID}, therefore, protons are
expected to be more suppressed than pions in $R_{AA}$ or $R_{CP}$
measurement. Experimentally, protons and pions show similar
magnitudes of suppression in $R_{CP}$~\cite{starAuAuPID}, which is
contradictory to the color charge dependence of energy loss
picture. Intrigued by our data, several calculations are proposed.
One of those is jet conversion mechanism~\cite{weiliu:07}, in
which, jet can change flavor or color charge after interaction
with the medium. Including both jet energy loss and conversion in
the expanding medium, the calculation results in a net quark to
gluon jet conversion. This leads to a better agreement with data.
With a much larger jet conversion cross section compared to that
in the Leading Order calculation, the proton and pion suppression
magnitudes are similar. The same idea are used to predict strange
hadron $R_{AA}$. Using the same factor scaling the leading order
(LO) QCD calculations, kaons are predicted to be less suppressed
with jet conversion than pions since the initially produced hard
strange quarks are much fewer than the strange quarks in a hot,
dense medium~\cite{Fries:08}. The interaction between the
initially produced light quark (gluon) and the medium will lead to
more strangeness production at high $p_T$ at RHIC energies
compared to the case without jet conversion. Alternatively,
enhanced parton splitting in the medium will also lead to a change
of the jet hadron chemical composition in Au+Au collisions
compared to that in p+p collisions~\cite{urs:07}. On the other
hand, non-photonic electrons, which are from heavy flavor charm
and bottom decay, show a similar magnitude of suppression as light
hadrons~\cite{starelectron}. The pQCD calculations including
collisional and radiative energy loss show a systematically higher
$R_{AA}$ value than experimental data~\cite{wicks,bdmps}. Further
calculations indicate that with charm contribution only,
non-photonic electrons are expected to reproduce the
data~\cite{bdmps}. In this proceedings, recent results on strange
hadron and heavy flavor at high $p_T$ from STAR are presented to
further understand energy loss mechanisms in the medium.

\section{Strange hadron measurements in p+p and Au+Au collisions}

\subsection{Constrain Fragmentation Functions (FFs)}
\begin{figure}[h] \centerline{
\includegraphics[width=0.55\textwidth]
{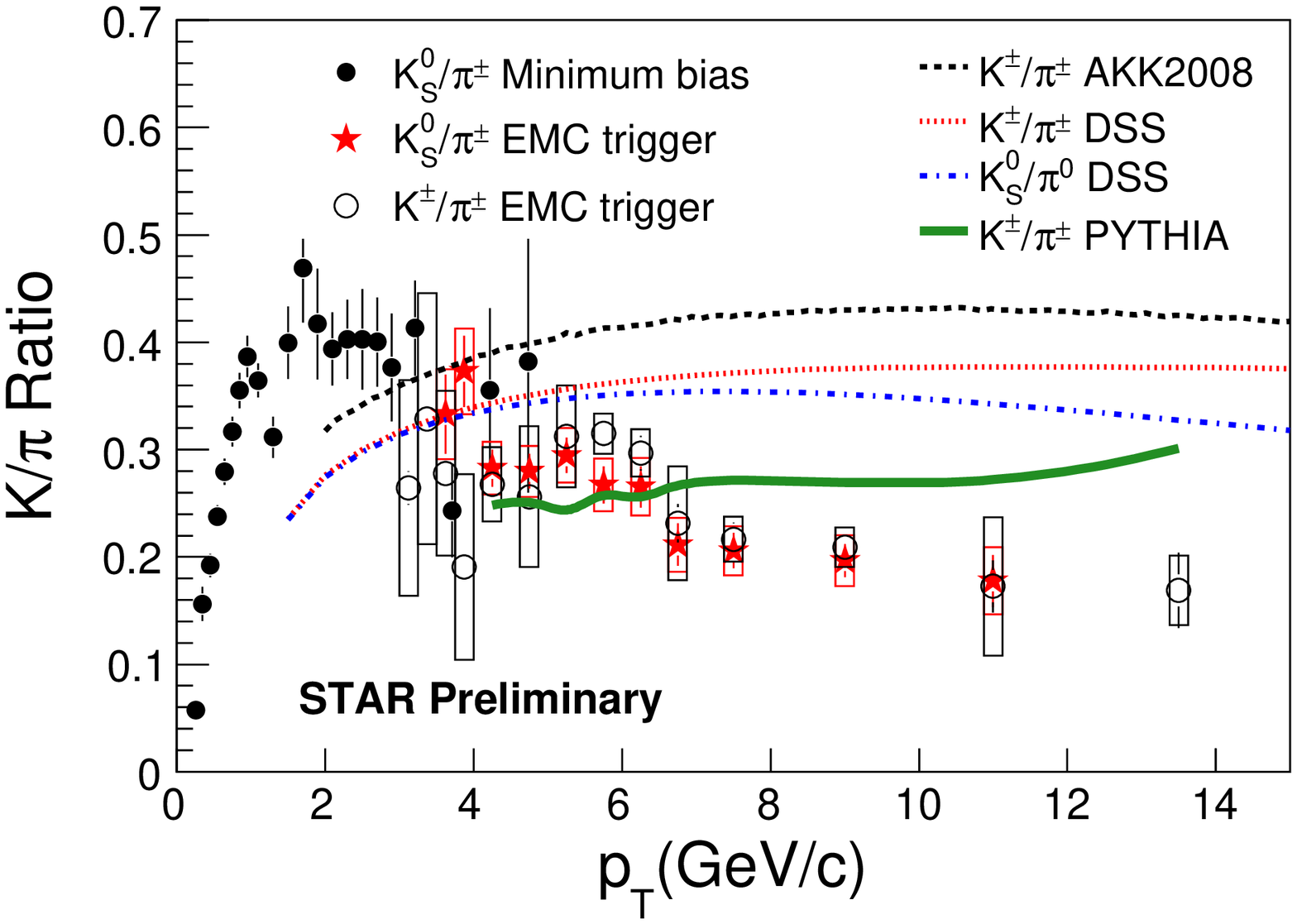}\includegraphics
[width=0.35\textwidth]{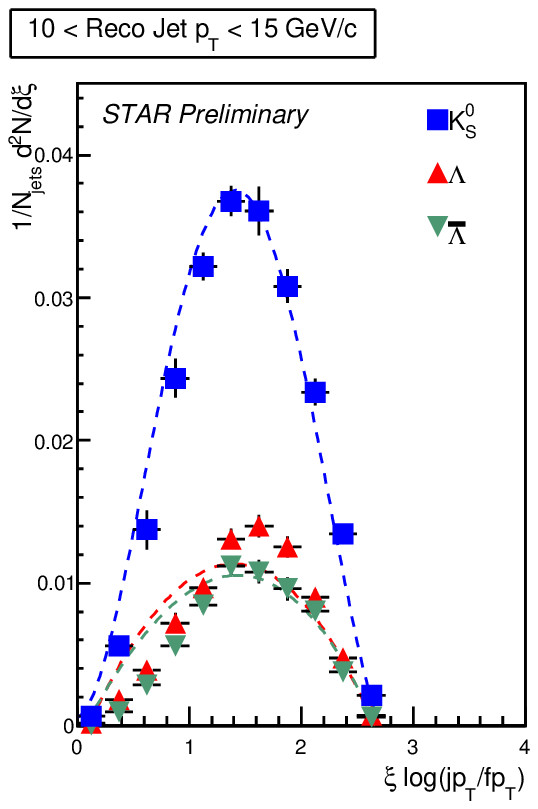}} \caption{(in color on line)
(left) The $K/\pi$ ratios in p+p collisions at 200 GeV. Shown as
different lines are calculations from NLO pQCD calculations with
AKK and DSS FFs and LO PYTHIA calculations. (right) The FF
measurements for $K_S^{0}$, $\Lambda$ and $\bar{\Lambda}$ in 200
GeV p+p collisions. Also shown are those from PYTHIA calculations
as dashed lines. The double counting correction was done while
missing energy, energy resolution and tracking efficiency have not
been corrected for.} \label{Figure1}
\end{figure}

To understand the medium properties using hard probes, it is
necessary to understand particle production in p+p collisions.
Using STAR Barrel Electron-magnetic Calorimeter (BEMC) triggered
events, identified particles such as $\pi^{\pm}$, $K^{\pm}$,
$p(\bar{p})$, $K_S^{0}$ and $\rho^{0}$ can be measured up to
$p_{T}\!=\!15$ GeV/$c$ in p+p collisions at $\sqrt{s_{NN}}$ = 200
GeV at mid-rapidity $|y|\!<0.5$. Figure~\ref{Figure1} (left) shows
the $K/\pi$ ratios measured in 200 GeV p+p collisions together
with those from next-to-leading order (NLO) pQCD calculations with
different FFs. In the overlapping $p_T$ region, the spectra of
$K^{\pm}$ and $K_S^{0}$ are consistent. The NLO pQCD calculations
with AKK~\cite{AKK} and DSS~\cite{DSS} FFs do not reproduce the
$K/\pi$ ratio while the LO PYTHIA calculations lead to a better
agreement with data. Not shown in this figure, but shown
in~\cite{yichunxu}, the NLO pQCD calculations with AKK and DSS FFs
can not reproduce the $p(\bar{p})/\pi$ ratios at high $p_T$ either
while the LO PYTHIA calculations lead to a better agreement with
data also. The identified particle spectra measurements of
$\pi^{\pm}$, $K$ and $p(\bar{p})$ at STAR should be able to
further constrain light flavor separated FFs and gluon FFs.

Taking advantage of the large acceptance of the STAR TPC, we can
measure the FFs of identified particles directly. Shown in
Fig.~\ref{Figure1} (right) are the FF measurements for $K_S^{0}$,
$\Lambda$ and $\bar{\Lambda}$ in 200 GeV p+p collisions with jet
energy of $10\!<\!p_T\!<\!15$ GeV/$c$. The details of the analysis
can be found at~\cite{timmins}. Interestingly, PYTHIA can
reproduce the $K_S^{0}$ FF reasonably well but shows deviation
from $\Lambda$ and $\bar{\Lambda}$ FF measurements.

\subsection{Jet hadro-chemistry change from p+p to Au+Au collisions}
\begin{figure}
\includegraphics*[keepaspectratio,scale=0.7]{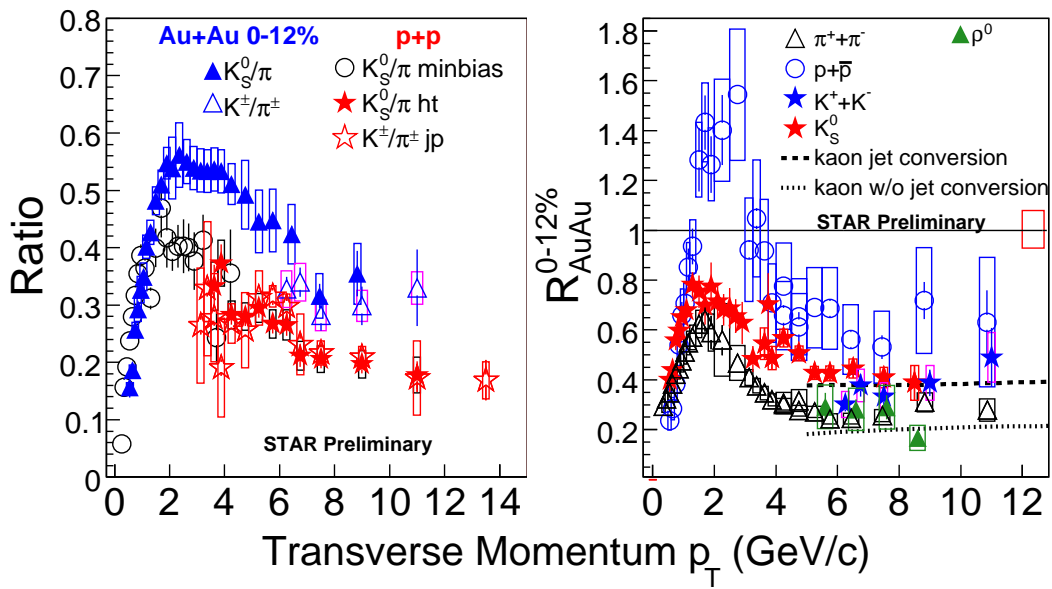}
\caption{(left) The $K^{\pm}(K_{S}^{0})/\pi^{\pm}$ ratios as a
function of $p_T$ in p+p and 0-12\% central Au+Au collisions.
(right) Nuclear modification factors of $\pi$, K, $p(\bar{p})$ and
$\rho$ in Au+Au collisions. The bars and boxes represent
statistical and systematic uncertainties.} \label{Figure2}
\end{figure}

Figure~\ref{Figure2} (left) shows the
$K^{\pm}(K_{S}^{0})/\pi^{\pm}$ ratios as a function of $p_T$ in
p+p and 0-12\% central Au+Au collisions. At $p_T\!>6$ GeV/$c$, the
$K/\pi$ ratio in central Au+Au collisions is higher than that in
p+p collisions, indicating that the jet hadro-chemistry change
from elementary p+p to central Au+Au collisions.
Figure~\ref{Figure2} (right) shows nuclear modification factors of
$\pi$, K, $p(\bar{p})$ and $\rho$ in Au+Au collisions. The
$\pi^{+}+\pi^{-}$ and $p+\bar{p}$ spectra in central Au+Au
collisions are from ref.~\cite{starAuAuPID}. We observe that
$R_{AA}$($K^{+}+K^{-}$,$K_S^{0}$) is larger than
$R_{AA}$($\pi^{+}+\pi^{-}$). This is consistent with the
prediction of jet conversion in the hot dense medium, as shown by
the dashed line, the prediction presented in ref.~\cite{Fries:08}.
However, using the same parameters this jet conversion scenario
predicts similar $R_{AA}$ for protons and pions~\cite{weiliu:07},
while experimentally we observe $R_{AA}$($p+\bar{p}$) seems to be
larger than $R_{AA}$($\pi^{+}+\pi^{-}$). The full systematic
uncertainties for protons are under study. Recently, it was argued
that a higher twisted effect, enhanced parton splitting, can alone
lead to significant changes in the jet hadron chemical
composition~\cite{urs:07}. In this model, heavier hadrons become
more abundant relative to the case without enhanced parton
splitting mechanism. For a jet with 50 GeV of energy, the kaon to
pion ratio increases by a factor $\sim$ 50\% and the proton to
pion ratio by a factor $\sim$ 100\%~\cite{urs:07}. Together with
the jet conversion mechanism, this might be able to explain our
observations. However, a full comparison requires consideration of
quantitative modelling and calculations incorporating 3D hydro in
an expanding medium~\cite{renk:07} and proper light
flavor-separated quark and gluon fragmentation functions.
Experimentally, high $p_T$ strange hadron measurements in d+Au,
its centrality dependence of $R_{AA}$ in Au+Au and elliptic flow
$v_{2}$ measurements will shed more light on our understanding of
energy loss mechanisms.

\section{Heavy flavor measurements}

\subsection{e-h correlation to access B contribution to non-photonic electrons in p+p}
Using the azimuthal angle correlations between non-photonic
electrons and charged hadrons (e-h) and between non-photonic
electrons and $D^{0}$ ($e-D^{0}$), we measured bottom contribution
factor to non-photonic electrons~\cite{ehcorrelation}. It was
found that at $p_T\!>5$ GeV/$c$, bottom contribution is about
50\%. This together with non-photonic electron $R_{AA}$
measurements still challenge the pQCD energy loss model
calculations, indicating that collisional dissociation of heavy
mesons~\cite{vitev:07}, in-medium heavy resonance
diffusion~\cite{rapp:06}, and multi-body
mechanisms~\cite{liuko:06} might play an important role for heavy
quark interaction with the medium. Due to the large systematic
uncertainty of the current non-photonic electron measurements, it
is not yet possible to distinguish different mechanisms mentioned
above and/or determine their possible relative contributions. In
the future, measurements on direct topological reconstruction of
heavy favor hadron decays are crucial to understand precisely the
energy loss of heavy flavor~\cite{hft}.

\subsection{high $p_T$ $J/\psi$ measurements in p+p and A+A collisions}
The dissociation of quarkonia due to color screening in a
Quark-Gluon Plasma (QGP) is thought to be a classic signature of
de-confinement in relativistic heavy-ion
collisions~\cite{colorscreen}. Results at RHIC show that the
suppression of the $J/\psi$ as a function of centrality (the
number of participants) is similar to that observed at the SPS,
even though the energy density reached in collisions at RHIC is
significantly higher~\cite{Adare:2006ns,Abreu:2000xe}. Possible
mechanisms such as sequential suppression~\cite{satz_0512217},
$c\bar{c}$
recombination~\cite{BraunMunzinger:2000px,Grandchamp:2001pf,Gorenstein:2001xp,Thews:2000rj}
were proposed to explain this. Recent Lattice QCD calculations
indicate that direct $J/\psi$ is not dissociated in the medium
created at RHIC while the suppression observed for $J/\psi$ comes
from the dissociation of $\chi_{c}$ and
$\psi^{'}$~\cite{latticeqcd}. However, the direct $J/\psi$ might
be dissociated at RHIC at high $p_{T}$, which was predicted in the
hot wind dissociation picture, in which the AdS/CFT approach was
used and the dissociation temperature for $J/\psi$ was predicted
to decrease as a function of $p_{T}$~\cite{adscft}. The AdS/CFT
approach was applied to hydro framework and predicted that
$J/\psi$ $R_{AA}$ decreases versus $p_{T}$~\cite{hydro+hotWind}.

\begin{figure}
\includegraphics*[keepaspectratio,scale=0.7]{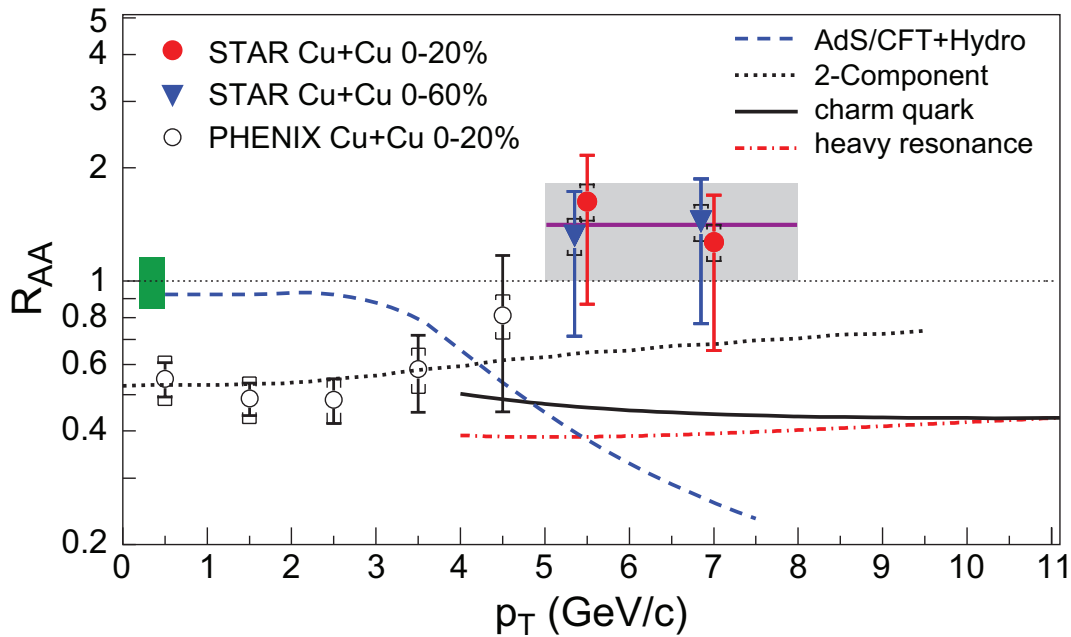}
\caption{$J/\psi$ $R_{AA}$ versus $p_{T}$. STAR data points have
statistical (bars) and systematic (caps) uncertainties. The box
about unity on the left shows $R_{AA}$ normalization uncertainty,
which is the quadrature sum of p+p normalization and binary
collision scaling uncertainties. The solid line and band show the
average and uncertainty of the two 0-20\% data points. The curves
are model calculations described in the text. The uncertainty band
of 10\% for the dotted curve is not shown.} \label{Figure3}
\end{figure}

Figure~\ref{Figure3} shows $J/\psi$ $R_{AA}$ as a function of
$p_{T}$ in 0-20\% and 0-60\% Cu+Cu collisions from
STAR~\cite{starjpsi} and 0-20\% Cu+Cu collisions from
PHENIX~\cite{phenixjpsi}. The average of two STAR 0-20\% data
points at high $p_T$ is $R_{AA}\!=\!1.4\! \!\pm\! \!0.4
(stat.)\!\pm\! \!0.2 (syst.)$. Compared to low $p_{T}$ PHENIX
measurements, our results indicate that $R_{AA}$ of $J/\psi$
increases from low $p_{T}$ to high $p_T$ at 97\% confidence level
(C.L.). The $R_{AA}$ of high $p_T$ $J/\psi$ is in contrast to
strong suppression for open charm~\cite{wicks,vitev:07,charmcucu},
indicating that $J/\psi$ might be dominantly produced through
color singlet configuration. However, even though there is
significant improvement from the next-next-to-leading order (NNLO)
pQCD calculations with color singlet model, the calculation still
fails to reproduce the high $p_T$ part~\cite{jpsicsNNLO}. The
$R_{AA}$ trend of $J/\psi$ is contradictory to
AdS/CFT+hydrodynamic calculations at 99\% C.L.. This might
indicate two things: 1) Cu+Cu system is not big enough so that the
calculation is not applicable. The larger system produced in Au+Au
collisions may be necessary to observe or exclude the effect
predicted by AdS/CFT; 2) the formation time effect for high $p_T$
$J/\psi$ is important since the AdS/CFT+hydrodynamic calculation
shown in Fig.~\ref{Figure3} requires that the $J/\psi$ be produced
as an on-shell $J/\psi$ fermion pair almost instantaneously at the
initial impact with no formation time. A calculation combining
effects of $J/\psi$ formation time, color screening, hadronic
phase dissociation, statistical $c\bar{c}$ coalescence and B meson
feed-down contribution can describe the
data~\cite{two_component_approach}. The calculation suggests a
slight increase in the $R_{AA}$ at higher $p_T$.

The excellent signal over background ratio of high $p_T$ $J/\psi$
enables to do two particle azimuthal angle correlations between
$J/\psi$ and charged hadrons ($J/\psi\!-\!h$) in p+p collisions.
From $J/\psi\!-\!h$ correlations, B meson contribution to
inclusive $J/\psi$ is obtained to be
$13\!\pm\!5$\%~\cite{starjpsi}. Even though there is no direct
measurement of B meson at RHIC, indirect measurements can help to
constrain B cross section in pQCD calculations through
$J/\psi\!-\!h$ and $e_{non-photonic}-h$ correlations, which
constrain B contribution factor to inclusive $J/\psi$ and
non-photonic electron yield respectively.

\subsection{$\Upsilon$ measurements at STAR}
Besides $J/\psi$, the $\Upsilon$ states are also ideal tools to
study the effect of color screening in hot and dense QCD matter
since its ground states and excited states melt at different
temperatures and all of them decay to
dileptons~\cite{satz_0512217}. Furthermore, since the $b\bar{b}$
cross section at RHIC energy is expected to be much smaller
compared to $c\bar{c}$  cross section from FONLL
calculations~\cite{vogt}, the recombination contribution from QGP
phase might be negligible to bottomonia production. This makes the
$\Upsilon$ even a better probe for studying the color screening
effect in QGP if sufficient statistics can be achieved
experimentally. STAR has measured
$\Upsilon\rightarrow\e^{+}\e^{-}$ in p+p, d+Au and Au+Au
collisions~\cite{reed}. The low material from run 8 significantly
reduced the background for $\Upsilon\rightarrow\e^{+}\e^{-}$
measurement. The $\Upsilon$ $R_{dAu}$ =
$0.98\!\pm\!0.32(stat.)\!\pm\!0.28(syst.)$, indicating that
$\Upsilon$ production follows $N_{bin}$ scaling in d+Au
collisions.

\subsection{Further measurements of heavy flavor} To further
understand the production mechanisms of quarkonia at high $p_T$
and medium properties, STAR will measure the following: nuclear
modification factors of $J/\psi$ from low to high $p_T$ in Au+Au
and d+Au collisions, $J/\psi$ $v_2$, forward and backward $J/\psi$
to address intrinsic charm contribution at large
$x_F$~\cite{perkins}, $J/\psi\!-\!h$ correlations to access
contribution factor from B, spin alignment of $J/\psi$, higher
charmonia states and different $\Upsilon$ states. The muons in
$\Upsilon\rightarrow\mu^{+}\mu^{-}$ do not suffer from
Bremsstrahlung radiation and with the Muon Telescope Detector
upgrade, we can cleanly separate the ground state from the excited
states even with the upgraded inner tracker with additional
material in the future~\cite{mtd}.

\section{Summary}
We report $K^{\pm}$ $p_T$ spectra at mid-rapidity ($|y|\!<\!0.5$)
up to 15 GeV/$c$, and neutral Kaon ($K_S^{0}$) $p_T$ spectra up to
12 GeV/$c$ using events triggered by the STAR BEMC from p+p
collisions at $\sqrt{s_{NN}}$ = 200 GeV. In p+p collisions, the
calculations from NLO pQCD models, fail to reproduce the $K^{\pm}$
spectra at high $p_T$. At $p_T\!>$ 6 GeV/$c$, we observe
$R_{AA}(K_{S}^{0}, K^{\pm},
p+\bar{p})\!>R_{AA}(\pi^{+}+\pi^{-})\!\sim\!
R_{AA}(\rho^{0})\!\sim\!R_{AA}\!(e_{non-photonic})$. This
challenges pQCD energy loss model calculations. High $p_T$
$J/\psi$ invariant cross sections in p+p and Cu+Cu collisions were
measured up to 14 GeV/$c$. The $R_{AA}$ of $J/\psi$ at $p_T\!>\!5$
GeV/$c$ is consistent with unity. These measurements together with
$\Upsilon$ results can further help to understand quarkonia
production mechanisms and medium properties. In the future, STAR
heavy flavor measurement will significantly benefit from detector
upgrade of Heavy Flavor Tracker~\cite{hft} and possible Muon
Telescope Detector~\cite{mtd}.

\end{document}